\documentclass[seceq]{ptptex}

\usepackage{epsfig}
\usepackage{rotating}
\usepackage{amsmath}
\usepackage{psfrag}

\newcommand{\vcb}{|V_{cb}|}
\newcommand{\vtd}{|V_{td}|}

\newcommand{\vts}{|V_{ts}|}
\newcommand{\vus}{|V_{us}|}

\newcommand{\ord}{{\cal O}}

\def\kpn{K^+\rightarrow\pi^+\nu\bar\nu}

\def\klpn{K_{\rm L}\rightarrow\pi^0\nu\bar\nu}

\newcommand{\tev}{\, {\rm TeV}}
\newcommand{\gev}{\, {\rm GeV}}
\newcommand{\mev}{\, {\rm MeV}}

\newcommand{\be}{\begin{equation}}
\newcommand{\ee}{\end{equation}}

\newcommand{\mw}{M_{\rm W}}

\title{
Testing the CKM Picture of Flavour and CP Violation in Rare K and B Decays and
Particle-Antiparticle Mixing
}

\author{
Andrzej J. \textsc{Buras}%
}

\inst{
Technical University Munich, Physics Department, D-85748 Garching, Germany,\\
 TUM Institute for Advanced Study, D-80333 M\"unchen, Germany
}

\abst{
We summarize briefly the CKM picture of flavour and CP violation that 
governs the models with minimal flavour violation (MFV). We then describe
how this framework can be efficiently tested  through
particle-antiparticle mixing and rare $K$ and $B$ decays.
In particular we provide a list of theoretically clean tests that the
simplest version of the MFV framework, the constrained MFV hypothesis,
has to face in the coming years.
Finally we offer a brief look at the most popular SM extensions that go
beyond the CKM framework like the general MSSM, Little Higgs model with 
T-parity and
Randall-Sundrum models with bulk fermions.
}

\begin{document}

\maketitle

\section{Preface}
It is a great honour to be able to contribute to this volume that celebrates
the 2008 Nobel Prize in Physics awarded to Kobayashi and Maskawa for their
seminal 1973 paper \cite{Kobayashi:1973fv} on  flavour and CP violation in 
the Standard Model (SM). 
However, 
I would like to emphasize that the recognition of this field by the Nobel 
Committee is not only the success of these two renowned Japanese physicists but
also a great success of Nicola Cabibbo whose seminal paper of 1963 
\cite{Cabibbo:1963yz} had
a tremendous impact on the field of flavour violation. Therefore this article
pays  tribute also to his work and I hope, together with many of my 
colleagues, that one day he will be awarded the Nobel Prize as well.

\section{Introduction}
The understanding of flavour dynamics is one of the most important goals of 
elementary particle physics. Because this understanding will likely come 
from very short distance scales, the loop induced processes like flavour 
changing neutral current (FCNC) transitions will for some time continue to 
play the crucial role in achieving this goal. They can be best studied in 
$K$ and $B$ decays but  $D$ decays and hyperon decays can also offer 
useful information in this respect. There is also a good chance that we
will learn a lot from FCNC processes in the top quark sector once the LHC
will start producing results.

Within the Standard Model,  FCNC processes are governed by
\begin{itemize}
\item
the unitary Cabibbo-Kobayashi-Maskawa (CKM) matrix 
\cite{Cabibbo:1963yz,Kobayashi:1973fv}
that parametrizes the weak charged current interactions of quarks,
\item
the Glashow-Iliopoulos-Maiani (GIM) mechanism \cite{Glashow:1970gm} 
that forbidds 
the appearance of FCNC 
processes at the tree level with the size of its violation at the one loop 
level depending sensitively on the CKM parameters and the masses of exchanged
particles,
\item
the asymptotic freedom of QCD \cite{Gross:1973id,Politzer:1973fx} that 
allows to calculate the 
impact of strong 
interactions on weak decays at sufficiently short distance scales within the 
framework of renormalization group improved perturbation theory,
\item
the operator product expansion (OPE) \cite{Wilson:1969zs} with
local operators having a specific Dirac structure and their matrix elements 
calculated by means of non--perturbative methods or in certain 
cases extracted from experimental data on tree level decays with the help
of flavour symmetries.
\end{itemize}

The present data on rare and CP violating $K$ and $B$ decays are consistent 
with this structure but as flavour physics enters only now a precision era and
many branching ratios  still have to be
measured, it is to be seen whether some modifications of this picture will
be required in the future when the data improve.

In order to appreciate the simplicity of the structure of FCNC processes 
within the SM, let us realize that although the CKM matrix was introduced
in connection with charged current interactions of quarks, 
its departure from the unit matrix is the origin of all flavour violating 
and CP-violating transitions in this model. Out there, at very short 
distance scales, the picture could still be very different. In particular, 
new complex phases could be present in both charged and neutral current 
interactions, the GIM mechanism could be violated already at the tree level, 
the number of parameters describing flavour violations could be significantly
larger than the four parameters 
present in the CKM matrix and the number of operators 
governing the decays could also be larger. We know all this through 
extensive studies of complicated extensions of the SM to which we will 
return briefly at the end of this writing.

In view of  New Physics (NP) waiting for us at very short distance scales
the SM is considered these
 days only as a low energy effective quantum field theory, based on the
 spontaneously broken gauge symmetry
\begin{equation}
SU(3)_C\otimes SU(2)_L\otimes U(1)_Y \to SU(3)_C \otimes U(1)_Q,
\end{equation}
that  describes low energy phenomena  in terms of 
28 parameters. The latter have to be determined from experiment. 
Two of these 
parameters $(\alpha_{\rm QCD},~\theta_{\rm QCD})$ are related to strong 
interactions and four to the electroweak gauge boson and Higgs sector.
 The remaining 22 parameters reside in the flavour sector: six quark masses, 
 six lepton masses, four  parameters of the CKM matrix 
\cite{Cabibbo:1963yz,Kobayashi:1973fv}
and six parameters of the PMNS
 matrix \cite{Pontecorvo:1957cp,Maki:1962mu}.

At first sight it would appear that while the success of the SM in 
describing the data in the strong interaction sector and electroweak gauge
 boson sector is very profound, the corresponding success in the flavour
 sector is rather obvious in view of so many free parameters. 
Yet in the case 
of the CKM picture of flavour changing interactions in the quark sector, 
combined with the GIM
 mechanism  \cite{Glashow:1970gm}  that governs FCNC
processes in the 
SM, such a view would totally misrepresent the facts.
Indeed, once all quark and lepton masses are determined, there are only the 
four
 free parameters of the CKM matrix to our disposal and in terms of them all 
existing data in the quark flavour sector can be properly described within
 experimental and theoretical uncertainties. Indeed, bearing in mind a few
 hints for the departures from the CKM picture of flavour and CP violation,
 to which we will return
 later on,
\begin{itemize}
\item
all leading decays of $K$, $D$, $B_d^0$ and $B_s^0$ mesons, that have been 
measured, are correctly described,
\item
suppressed transitions in the SM, like $K^0-\bar K^0$ mixing, 
$B_d^0-\bar B_d^0$ mixing and $B^0_s-\bar B^0_s$ mixing have 
not only been found at the suppressed level, but even at the predicted order
 of magnitude and in fact even better than that,
\item
CP-violating observables in $K_L$, $K^\pm$, $B_d^0$ and $B^\pm$ decays agree 
 well with the existing data and
\item
the best measured semi-rare (radiative) B-decays: $B\to X_s\gamma$, 
$B\to X_s l^+l^-$ and $B_d\to K^*\gamma$  all turned out to have 
branching ratios close to the SM predictions.
\end{itemize}
{
\begin{table}[ht]
\renewcommand{\arraystretch}{1}\setlength{\arraycolsep}{1pt}
\center{\begin{tabular}{|c|c|c|c|c|c|}
\hline
       &$B_s\to \mu^+\mu^-$ & $\klpn$ & $K_L\to\mu e$ & $\mu\to e\gamma$ &
               $d_n$ \\
\hline
 SM  & $3\cdot 10^{-9}$ & $3\cdot 10^{-11}$ & $10^{-40}$ & $10^{-54}$ &
                $ 10^{-32}$ e cm.    \\
 Exp Bound  &  $4\cdot 10^{-8}$  & $6\cdot 10^{-8}$ &  $10^{-12}$ &
     $10^{-11}$ & $5\cdot 10^{-26}$ e cm. \\
 \hline 
\end{tabular}  }
\caption {Approximate SM  values and experimental upper bounds for 
 selected branching ratios and the neutron electric dipole moment $d_n$.
}
\label{tab:bounds}
\renewcommand{\arraystretch}{1.0}
\end{table}
}
But this is not the whole story, as many very strongly suppressed
 branching ratios within the SM are also consistent with experiment: 
the corresponding decays have not been observed yet. Examples are collected
in Table~\ref{tab:bounds}, where we compare approximate SM values with
the experimental upper bounds. Clearly there is still a lot of room for
NP contributions.

However, one of the very suppressed decays has been seen. It is
 $\kpn$  which in the SM is predicted to have the branching ratio
$Br(\kpn)=(8.5\pm 0.7)\cdot 10^{-11}$ \,\cite{Brod:2008ss}. 
Seven events have been found 
implying 
$Br(\kpn)=(17\pm 11)\cdot 10^{-11}$ \,\cite{Artamonov:2008qb}, 
on the high side but still
consistent with the SM value. 

In spite of all these successes the situation is certainly not satisfactory. 
Indeed,
\begin{itemize}
\item
the neutral Higgs boson has not been found yet,
\item
the Higgs mass $m_H$ is plagued by quadratic divergences present in the 
one-loop contributions to the Higgs propagator with internal top quark, gauge boson and 
Higgs exchanges. Within the SM there is no protective symmetry that would 
keep $m_H=\ord(v_{\rm ew})$ and if we want to assure this in the presence of a
 cut-off as high as $\Lambda_{\rm Planck}$, a fantastic 
fine tuning of SM parameters has to be made, which is obviously very 
unnatural,
\item
the hierarchical structures of quark and lepton masses and of their flavour 
violating interactions parametrized by the CKM and PMNS matrices remain 
a mystery, which at least from my point of view has not been satisfactorily 
uncovered in spite of intensive efforts during the last 30 years. But there
are some interesting advances which we will mention briefly  later on.
\end{itemize}

There are clearly other issues like the quantization of electric charge, 
the number of 
quark and lepton generations, the number of space dimensions, 
the baryon-antibaryon asymmetry in the universe, dark 
matter and dark energy, but I do not have space to  address them here. 
Similarly, I do not have space to address in detail 
the tests of various extensions
of the SM like general supersymmetric models, Little Higgs models and 
Randall-Sundrum models.
 What I would like to do primarly 
here, in view of
the recent Nobel Prize, is to summarize briefly the present status of the
CKM picture of flavour and CP violation that goes beyond the SM itself and
encompases all models with constrained minimal flavour violation (CMFV) and
more general models with MFV. In particular we will summarize stringent tests
through rare $K$ and $B$ decays that hopefully will tell us one day whether 
the CKM
picture is indeed the whole story.

The next pages recall briefly the CKM matrix and the related unitarity
triangle (UT) and describe the concepts of CMFV and MFV. Subsequently we
will discuss the most stringent tests of these frameworks. We next
collect signals of NP beyond CMFV and MFV that seem to be present in 
a number of experimental results. Before closing with an outlook we will
briefly summarize the most prominent non-MFV extensions of the SM and their
implications for FCNC processes.

\section{CKM Matrix and the Unitarity Triangle }
The unitary CKM matrix connects  the {\it weak
eigenstates} $(d^\prime,s^\prime,b^\prime)$ and 
 the corresponding {\it mass eigenstates} $d,s,b$:
\begin{equation}\label{2.67}
\left(\begin{array}{c}
d^\prime \\ s^\prime \\ b^\prime
\end{array}\right)=
\left(\begin{array}{ccc}
V_{ud}&V_{us}&V_{ub}\\
V_{cd}&V_{cs}&V_{cb}\\
V_{td}&V_{ts}&V_{tb}
\end{array}\right)
\left(\begin{array}{c}
d \\ s \\ b
\end{array}\right)\equiv\hat V_{\rm CKM}\left(\begin{array}{c}
d \\ s \\ b
\end{array}\right).
\end{equation}

Many parametrizations of the CKM
matrix have been proposed in the literature \cite{Fritzsch:1999ee}. While the so called 
standard parametrization \cite{Chau:1984fp} 
\begin{equation}\label{2.72}
\hat V_{\rm CKM}=
\left(\begin{array}{ccc}
c_{12}c_{13}&s_{12}c_{13}&s_{13}e^{-i\delta}\\ -s_{12}c_{23}
-c_{12}s_{23}s_{13}e^{i\delta}&c_{12}c_{23}-s_{12}s_{23}s_{13}e^{i\delta}&
s_{23}c_{13}\\ s_{12}s_{23}-c_{12}c_{23}s_{13}e^{i\delta}&-s_{23}c_{12}
-s_{12}c_{23}s_{13}e^{i\delta}&c_{23}c_{13}
\end{array}\right)\,,
\end{equation}
with
$c_{ij}=\cos\theta_{ij}$ and $s_{ij}=\sin\theta_{ij}$ 
($i,j=1,2,3$) and the complex phase $\delta$ necessary for {\rm CP} violation,
should be recommended  
for any numerical 
analysis, the Wolfenstein parametrization 
\cite{Wolfenstein:1983yz} and its more accurate generalization \cite{Buras:1994ec} to higher
orders in $\lambda=\vus$
are more  
transparent than the standard parametrization and allow a fast estimate of
different contributions to a given decay amplitude. 

To this end we make the following change of variables in
the standard parametrization (\ref{2.72}) 
\cite{Buras:1994ec}
\begin{equation}\label{2.77} 
s_{12}=\lambda\,,
\qquad
s_{23}=A \lambda^2\,,
\qquad
s_{13} e^{-i\delta}=A \lambda^3 (\varrho-i \eta)
\end{equation}
where
\begin{equation}\label{2.76}
\lambda, \qquad A, \qquad \varrho, \qquad \eta \, 
\end{equation}
are the Wolfenstein parameters
with $\lambda\approx 0.225$ being the expansion parameter. We find then 
\begin{equation}\label{f1}
V_{ud}=1-\frac{1}{2}\lambda^2-\frac{1}{8}\lambda^4, \qquad
V_{cs}= 1-\frac{1}{2}\lambda^2-\frac{1}{8}\lambda^4(1+4 A^2),
\end{equation}
\begin{equation}
V_{tb}=1-\frac{1}{2} A^2\lambda^4, \qquad
V_{cd}=-\lambda+\frac{1}{2} A^2\lambda^5 [1-2 (\varrho+i \eta)],
\end{equation}
\begin{equation}\label{VUS}
V_{us}=\lambda+\ord(\lambda^7),\qquad 
V_{ub}=A \lambda^3 (\varrho-i \eta), \qquad 
V_{cb}=A\lambda^2+\ord(\lambda^8),
\end{equation}
\begin{equation}\label{2.83d}
 V_{ts}= -A\lambda^2+\frac{1}{2}A\lambda^4[1-2 (\varrho+i\eta)],
\qquad V_{td}=A\lambda^3(1-\bar\varrho-i\bar\eta),
\end{equation}
where terms 
$\ord(\lambda^6)$ and higher order terms have been neglected.
A non-vanishing $\eta$ is responsible for CP violation in the MFV models.
 It plays 
the role of $\delta$ in the standard parametrization.
Finally, the barred variables in (\ref{2.83d}) are given by
\cite{Buras:1994ec}
\begin{equation}\label{2.88d}
\bar\varrho=\varrho (1-\frac{\lambda^2}{2}),
\qquad
\bar\eta=\eta (1-\frac{\lambda^2}{2}).
\end{equation}

Now, the unitarity of the CKM-matrix implies various relations between its
elements. In particular, we have
\begin{equation}\label{2.87h}
V_{ud}^{}V_{ub}^* + V_{cd}^{}V_{cb}^* + V_{td}^{}V_{tb}^* =0.
\end{equation}
The relation (\ref{2.87h})  can be
represented as a ``unitarity'' triangle in the complex 
$(\bar\varrho,\bar\eta)$ plane. 
One can construct five additional unitarity triangles 
\cite{Aleksan:1994if,Bigi:1999hr} 
corresponding to other unitarity relations.

Noting that to an excellent accuracy $V_{cd}^{}V_{cb}^*$ is real with
$| V_{cd}^{}V_{cb}^*|=A\lambda^3+\ord(\lambda^7)$ and
rescaling all terms in (\ref{2.87h}) by $A \lambda^3$ 
we indeed find that the relation (\ref{2.87h}) can be represented 
as the triangle 
in the complex $(\bar\varrho,\bar\eta)$ plane 
as shown in fig.~\ref{fig:utriangle}. Let us collect some 
useful formulae related 
to this triangle:

\begin{figure}
\begin{center}
\includegraphics[width=3in]{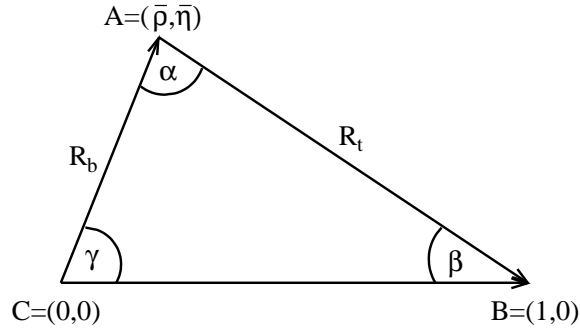}
\end{center}
\caption{\label{fig:utriangle}Unitarity Triangle. In Japanese $\alpha=\phi_2$,
$\beta=\phi_1$ and $\gamma=\phi_3$.}
\end{figure}

\begin{itemize}
\item
We can express $\sin(2\phi_i$), $\phi_i=
\beta,\alpha,\gamma$, in terms of $(\bar\varrho,\bar\eta)$. In particular:
\begin{equation}\label{2.90}
\sin(2\beta)=\frac{2\bar\eta(1-\bar\varrho)}{(1-\bar\varrho)^2 + \bar\eta^2}.
\end{equation}
\item
The lengths $CA$ and $BA$ are given (respectively) by 
\begin{equation}\label{2.94}
R_b \equiv \frac{| V_{ud}^{}V^*_{ub}|}{| V_{cd}^{}V^*_{cb}|}
= \sqrt{\bar\varrho^2 +\bar\eta^2}
= (1-\frac{\lambda^2}{2})\frac{1}{\lambda}
\left| \frac{V_{ub}}{V_{cb}} \right|,
\end{equation}
\begin{equation}\label{2.95}
R_t \equiv \frac{| V_{td}^{}V^*_{tb}|}{| V_{cd}^{}V^*_{cb}|} =
 \sqrt{(1-\bar\varrho)^2 +\bar\eta^2}
=\frac{1}{\lambda} \left| \frac{V_{td}}{V_{cb}} \right|.
\end{equation}
\item
The angles $\beta$ and $\gamma=\delta$ of the unitarity triangle 
are related
directly to the complex phases of the CKM elements $V_{td}$ and
$V_{ub}$, respectively, through
\begin{equation}\label{e417}
V_{td}=|V_{td}|e^{-i\beta},\quad V_{ub}=|V_{ub}|e^{-i\gamma}.
\end{equation}
\item
The unitarity relation (\ref{2.87h}) can be rewritten as
\begin{equation}\label{RbRt}
R_b e^{i\gamma} +R_t e^{-i\beta}=1~.
\end{equation}
\item
The angle $\alpha$ can be obtained through the relation
\begin{equation}\label{e419}
\alpha+\beta+\gamma=180^\circ~.
\end{equation}
\end{itemize}

Formula (\ref{RbRt}) shows transparently that the knowledge of
$(R_t,\beta)$ allows to determine $(R_b,\gamma)$ through 
\begin{equation}\label{VUBG}
R_b=\sqrt{1+R_t^2-2 R_t\cos\beta}~,\qquad
\cot\gamma=\frac{1-R_t\cos\beta}{R_t\sin\beta}.
\end{equation}
Similarly, $(R_t,\beta)$ can be expressed through $(R_b,\gamma)$ by
\begin{equation}\label{VTDG}
R_t=\sqrt{1+R_b^2-2 R_b\cos\gamma}~,\qquad
\cot\beta=\frac{1-R_b\cos\gamma}{R_b\sin\gamma}.
\end{equation}
These relations are remarkable. They imply that the knowledge 
of the coupling $V_{td}$ between $t$ and $d$ quarks allows to deduce the 
strength of the corresponding coupling $V_{ub}$ between $u$ and $b$ quarks 
and vice versa.

The triangle depicted in fig. \ref{fig:utriangle}, together with $|V_{us}|$ 
and $\vcb$ gives the full description of the CKM matrix. 
Looking at the expressions for $R_b$ and $R_t$, we observe that within
the MFV models the measurements of four CP
{\it conserving } decays sensitive to $|V_{us}|$, $|V_{ub}|$,   
$|V_{cb}|$ and $|V_{td}|$ can tell us whether CP violation
($\bar\eta \not= 0$ or $\gamma \not=0,\pi$) is present or not
 in the MFV models. 
This property is often used to determine
the angles of the unitarity triangle without the study of CP-violating
quantities. It constitues a very important test of the CMFV and MFV frameworks.

\section{Theory of CMFV }
\subsection{Master Formula}
As already stated at the beginning of this writing the physics at very
short distances could in principle deviate profoundly from the CKM picture
of flavour and CP violation. However, it is also possible that this picture
will dominantly describe all the data available in the future. This would
not imply that there is no NP beyond the SM but only that the flavour
and CP violation in the quark sector is governed even beyond the SM by the
CKM matrix or equivalently by the structure of quark Yukawa couplings. This is 
the MFV hypothesis. Let us
state this hypothesis in explicit terms by using the master formula for
weak decays that follows from the operator product expansion and 
renormalization group approach.

 The master formula in question reads \cite{Buras:2001pn}
\begin{equation}\label{master}
{\rm A(Decay)}=\sum_i B_i \eta^i_{\rm QCD}V^i_{\rm CKM} 
F_i(v),
\end{equation}
where $B_i$ are non-perturbative parameters representing hadronic matrix
elements of the contributing operators, $\eta_i^{\rm QCD}$ stand symbolically 
for the renormalization group QCD factors \cite{Buchalla:1995vs}, $V^i_{\rm CKM}$ denote 
the relevant combinations of the elements of the CKM matrix and finally 
$F_i(v)$
denote the loop functions that in most models result from box and penguin
diagrams but in some
models can also represent tree level diagrams if such diagrams contribute.
The variable $v$ collects all parameters in addition to $m_t$, in 
particular the set of new gauge 
couplings $g_i^{\rm NP}$, masses of new particles
$m_i^{\rm NP}$ and new flavour and CP violating couplings $V^{ij}_{\rm  NP}$.
It turns out to be useful to factor out $V^i_{\rm CKM}$ in all contributions
in order to see transparently the deviations from MFV.
In writing (\ref{master}) we did not show explicitly the internal charm 
contributions that cannot be neglected in certain $K$ decays but are usually
very small in $B$ decays.

Now, in the SM only a particular set of parameters $B_i$ is relevant, 
the functions $F_i$ are {\it real} and the  flavour and CP violating effects
 enter only through the CKM factors $V^i_{\rm CKM}$. This  implies that the
functions $F_i$ are universal with respect to flavour so that they are the
same in the $K$, $B_d$ and $B_s$ systems. Consequently a number of observables 
in these systems are strongly correlated in the SM.

 The simplest class of extensions of the SM are  models with 
constrained Minimal Flavour Violation (CMFV)
\cite{Buras:2000dm, Buras:2003jf,Blanke:2006ig}.
They are formulated as follows:
\begin{itemize}
\item
All flavour changing transitions are governed by the CKM matrix with the 
CKM phase being the only source of CP violation,
\item
The only relevant operators in the effective Hamiltonian below the weak scale
are those that are also relevant in the SM.
\end{itemize}
This implies that relatively to the SM  only the values of $F_i$ are 
modified but their universal character remains intact. Moreover,
 in cases where 
$F_i$ can be eliminated by taking certain combinations of observables,
universal correlations between these observables
 for this class of models result. We will list
them below.

The SM, the Two Higgs Doublet Model II with a moderate 
$\tan\beta$, the SM with one extra
universal flat dimension and the Littlest Higgs model without T parity are
prominent members of this class of models.

More generally \cite{Chivukula:1987py,Hall:1990ac}, as formulated elegantly 
with the help of global symmetries and 
the spurion technique \cite{D'Ambrosio:2002ex}, the operator structure 
in MFV models can differ from
the SM one if two Higgs doublets are present and bottom and top Yukawa 
couplings are of comparable size. A well known example is the MSSM with MFV
and large $\tan\beta$. In these models new parameters $B_i$ and 
$\eta_i^{\rm QCD}$, related to new operators enter the game but the functions
$F_i(v)$ still remain real quantities as in the CMFV
framework and do not involve any flavour violating parameters, so that
the CP and flavour violatiing effects are again governed by the CKM matrix.
However, the presence of new operators makes this approach less constraining
than the CMFV framework.

In the simplest non-MFV models, the basic operator structure of CMFV models
remains unchanged but the functions $F_i$ in addition to real SM contributions can
contain new flavour parameters and new complex phases. Consequently the
CKM matrix ceases to be the only source of flavour and CP violation.
The Littlest Higgs model with T-parity and 
the SM extended to four generations are prominent members of this class
of models. 

Finally, in most general non-MFV models, new operators (new $B_i$ parameters)
contribute and the functions $F_i$ in addition to real SM contributions can
contain new flavour parameters and new complex phases. Here the prominent
members are the general MSSM, Randall-Sundrum models and generally models
with FCNC transitions at the tree level. 
Also the NMFV framework \cite{Agashe:2005hk} can be classified here.

We postpone the discussion of these non-MFV models to the
last section and we will first concentrate on the properties of the 
CMFV models.

\subsection{Master Functions in the CMFV}
The CMFV  models 
can be formulated to a very good approximation 
in terms of 11 parameters \cite{Buras:2003jf}: four parameters of the CKM matrix and seven values 
of the real master functions $F_i(v)$ that enter the master formula 
(\ref{master}) and parametrize the short distance 
contributions. In a given CMFV model, the $F_i$ can be calculated in 
perturbation theory and are generally correlated with each other but 
in a model independent analysis they must be considered as free 
parameters. Explict calculations indicate that five or even only four  of these 
functions receive significant new physics contributions.

The master functions $F_i(v)$ originate from
various penguin and box diagrams. 
In order to find these master functions  we first express
the penguin vertices (including electroweak counter terms)
in terms of the functions
$C$ ($Z^0$ penguin), $D$ ($\gamma$ penguin), $E$ (gluon penguin), 
$D'$ ($\gamma$-magnetic penguin) and $E'$ (chromomagnetic 
penguin). Similarly we can define the box diagram function $S$ 
($\Delta F=2$ transitions), 
as well as $\Delta F=1$ box functions
$B^{\nu\bar\nu}$ and $B^{\mu\bar\mu}$ relevant for decays with 
${\nu\bar\nu}$  and ${\mu\bar\mu}$ in the final state, respectively.
The result of this exercise exists \cite{Buras:1998raa}.

While the $\Delta F=2$ box function $S$ and the penguin functions 
$E$, $D'$ and $E'$ are gauge independent, this is not the case for 
$C$, $D$ and the $\Delta F=1$ box diagram functions 
$B^{\nu\bar\nu}$ and $B^{\mu\bar\mu}$.
In  phenomenological applications it is more 
convenient to work with gauge independent functions \cite{Buchalla:1990qz}
given by
\begin{equation}\label{XYZ} 
X(v)=C(v)+B^{\nu\bar\nu}(v),\qquad  
Y(v)  =C(v)+B^{\mu\bar\mu}(v), \qquad
Z(v)  =C(v)+\frac{1}{4}D(v).
\end{equation}
Indeed, the box diagrams in the CMFV framework 
have the Dirac structure $(V-A)\otimes (V-A)$, 
the $Z^0$ penguin diagram has the $(V-A)\otimes(V-A)$ and 
$(V-A)\otimes V$ components and the $\gamma$ penguin is pure
$(V-A)\otimes V$.
The $X$ and $Y$ functions correspond then to linear combinations of the 
$(V-A)\otimes(V-A)$ component 
of the $Z^0$ penguin diagram and box diagrams with final state 
quarks and leptons 
having weak isospin $T_3=1/2$ and $T_3=-1/2$, respectively. The $Z$ 
function corresponds 
to the linear combination of  the $(V-A)\otimes V$ component 
of the $Z^0$ penguin diagram and the $\gamma$ penguin.

Then the set of seven gauge independent master functions which govern
the FCNC processes in the CMFV models is given by
\be\label{masterf}
S(v),~X(v),~Y(v),~Z(v),~E(v),~ D'(v),~ E'(v)~.
\ee

Generally, several master functions contribute to a given decay,
although decays exist which depend only on a single function.
We have the following correspondence between the most interesting FCNC
processes and the master functions in the CMFV models:
\begin{center}
\begin{tabular}{lcl}
$K^0-\bar K^0$-mixing ($\varepsilon_K$) 
&\qquad\qquad& $S(v)$ \\
$B_{d,s}^0-\bar B_{d,s}^0$-mixing ($\Delta M_{s,d}$) 
&\qquad\qquad& $S(v)$ \\
$K \to \pi \nu \bar\nu$, $B \to X_{d,s} \nu \bar\nu$ 
&\qquad\qquad& $X(v)$ \\
$K_{\rm L}\to \mu \bar\mu$, $B_{d,s} \to l\bar l$ &\qquad\qquad& $Y(v)$ \\
$K_{\rm L} \to \pi^0 e^+ e^-$ &\qquad\qquad& $Y(v)$, $Z(v)$, 
$E(v)$ \\
$\varepsilon'$ &\qquad\qquad& $X(v)$, $Y(v)$, $Z(v)$,
$E(v)$ \\
$B \to X_s \gamma$ &\qquad\qquad& $D'(v)$, $E'(v)$ \\
$B \to X_s~{\rm gluon}$ &\qquad\qquad& $E'(v)$ \\
$B \to X_s l^+ l^-$ &\qquad\qquad&
$Y(v)$, $Z(v)$, $E(v)$, $D'(v)$, $E'(v)$
\end{tabular}
\vspace*{0.6cm}
\end{center}

This table means that the observables like branching ratios, mass differences
$\Delta M_{d,s}$ in $B_{d,s}^0-\bar B_{d,s}^0$-mixing and the CP violation 
parameters $\varepsilon$ und $\varepsilon'$, all can be to a very good 
approximation entirely expressed in
terms of the corresponding master functions and the relevant CKM factors. The
remaining entries in the relevant formulae for these observables are the 
renormalization group QCD factors $\eta_i$ and the
non-perturbative parameters $B_i$ that can be calculated by lattice methods in 
the SM or in certain cases extracted from experiment.

We know from the study of FCNC processes that not all master functions are 
important in a given decay. In fact plausible arguments exist
 \cite{Buras:2003jf} that only
the functions
\be\label{rmasterf}
S(v),~C(v),~ D'(v),~ E'(v)
\ee
are significantly affected by NP contributions so that the remaining
entries in (\ref{masterf}) can be to first approximation replaced by
their SM values if one does not aim at high precision. In this 
approximation the CMFV models depend effectively on eight parameters.

\subsection{Model Independent Relations}
The simple structure of CMFV models allows to derive a number of relations 
between various observables that 
do not depend on the functions $F_i(v)$ and consequently are universal 
within this class of models. Let us list the most important relations.

{\bf 1.}
There exists a universal unitarity triangle (UUT) 
\cite{Buras:2000dm} common to all 
these models and the SM that can be constructed by using measurable 
quantities that depend on the CKM parameters but are not polluted by the 
new parameters present in the extensions of the SM. 
The UUT can be constructed, for instance, by using $\sin 2\beta$ from 
the mixing induced CP asymmetry
$S_{\psi K_S}$ and the ratio $\Delta M_s/\Delta M_d$. 
The relevant formulae are given in (\ref{RRt}) and (\ref{sin2b}).

{\bf 2.} Next we have
\begin{equation}\label{dmsdmd}
\frac{\Delta M_d}{\Delta M_s}=
\frac{m_{B_d}}{m_{B_s}}
\frac{\hat B_{d}}{\hat B_{s}}\frac{F^2_{B_d}}{F^2_{B_s}}
\left|\frac{V_{td}}{V_{ts}}\right|^2,
\end{equation}
\begin{equation}\label{bxnn}
\frac{Br(B\to X_d\nu\bar\nu)}{Br(B\to X_s\nu\bar\nu)}=
\left|\frac{V_{td}}{V_{ts}}\right|^2,
\end{equation}
\begin{equation}\label{bmumu}
\frac{Br(B_d\to\mu^+\mu^-)}{Br(B_s\to\mu^+\mu^-)}=
\frac{\tau({B_d})}{\tau({B_s})}\frac{m_{B_d}}{m_{B_s}}
\frac{F^2_{B_d}}{F^2_{B_s}}
\left|\frac{V_{td}}{V_{ts}}\right|^2
\end{equation}
that all can be used to determine $|V_{td}/V_{ts}|$ without the knowledge
of $F_r(v)$. In particular,
the relation (\ref{dmsdmd})  with the precisely measured mass
differences  $\Delta M_{d,s}$ offers a  powerful determination of the 
length of one side of the unitarity triangle, denoted usually by $R_t$.
One finds
\cite{Blanke:2006ig}
\be\label{RRt}
(R_t)_{\rm CMFV}
\approx 0.90~\left[\frac{\xi}{1.21}\right] 
\sqrt{\frac{17.8/\text{ps}}{\Delta M_s}} 
\sqrt{\frac{\Delta M_d}{0.507/\text{ps}}},
\ee
where 
\be\label{xi}
\xi = 
\frac{\sqrt{\hat B_{B_s}}F_{B_s} }{ \sqrt{\hat B_{B_d}}F_{B_d}}=1.21\pm 0.04,
\qquad \xi=1.258\pm0.033
\ee
as summarized  by Lubicz and Tarantino \cite{Lubicz:2008am} and by  the HPQCD collaboration 
\cite{Gamiz:2009ku}, 
respectively.
The expression (\ref{RRt}) and 
\be\label{sin2b}
(\sin 2\beta)_{\rm CMFV}=(\sin 2\beta)_{\psi K_S}
\ee
allow to construct the UUT.

{\bf 3.} Eliminating $|V_{td}/V_{ts}|$ from the three 
relations above allows 
to obtain three relations between observables that are universal within the
CMFV models. In particular 
from (\ref{dmsdmd}) and (\ref{bmumu}) one finds the first ``golden'' relation
of CMFV \cite{Buras:2003td} 
\begin{equation}\label{R1}
\frac{Br(B_{s}\to\mu\bar\mu)}{Br(B_{d}\to\mu\bar\mu)}
=\frac{\hat B_{d}}{\hat B_{s}}
\frac{\tau( B_{s})}{\tau( B_{d})} 
\frac{\Delta M_{s}}{\Delta M_{d}},
\end{equation}
that does not 
involve the decay constants $F_{B_q}$ and consequently contains 
substantially smaller hadronic uncertainties than the formulae considered 
above.
 Indeed the present lattice values from \cite{Lubicz:2008am} read
\be
F_{B_s} \sqrt{\hat B_{B_s}} = 270(30)\mev,\qquad
F_{B_d} \sqrt{\hat B_{B_d}} = 225(25)\mev,
\ee
while the HPQCD collaboration \cite{Gamiz:2009ku} finds similar values but smaller errors
 \be
F_{B_s} \sqrt{\hat B_{B_s}} = 266(18)\mev,\qquad
F_{B_d} \sqrt{\hat B_{B_d}} = 216(15)\mev.
\ee

Note that the simple relation in (\ref{R1}) involves
only measurable quantities except for the ratio $\hat B_{s}/\hat B_{d}$
that is known with respectable precision \cite{Lubicz:2008am}
\begin{equation}\label{BBB}
\frac{\hat B_{s}}{\hat B_{d}}=1.00\pm 0.03, \qquad
\hat B_{d}=1.22\pm0.12, \qquad \hat B_{s}=1.22\pm0.12~.
\end{equation}

The formulae given above imply three universal results  for CMFV models:
\be\label{R1A}
\frac{Br(B_{s}\to\mu^+\mu^-)}{Br(B_{d}\to\mu^+\mu^-)}
= 32.5\pm1.7
\ee
\be\label{Bnunu}
\frac{Br(B\to X_s\nu\bar\nu)}{Br(B\to X_d\nu\bar\nu)}
=\frac{\vts^2}{\vtd^2}
=\frac{m_{B_d}}{m_{B_s}}
\frac{1}{\xi^2}
\frac{\Delta M_s}{\Delta M_d}=22.8\pm2.2,
\end{equation}
\be\label{vtdvts}
\frac{\vtd}{\vts}=0.210\pm0.011~,
\ee
where we have used the most recent values for the relevant quantities 
\cite{Lubicz:2008am}. When $\xi$ from HPQCD in (\ref{xi}) is used, the 
central values in (\ref{Bnunu}) and (\ref{vtdvts}) are changed to $21.1$ and 
$0.218$, respectively.
The three numbers in (\ref{R1A})--(\ref{vtdvts}) are universal magic numbers of CMFV. Non-confirmation
of these numbers in future experiments would signal non-CMFV contributions.
Only the last number can be tested at present and as seen in (\ref{vtdvtse})
CMFV survives this test.

{\bf 4.} Next the relation (\ref{R1}) allows to predict the branching ratios 
for $B_{s,d}\to\mu^+\mu^-$ within the 
SM and any CMFV model with much higher accuracy than it is 
possible without $\Delta M_{s,d}$. In the SM one has \cite{Buras:2003td}
\be\label{R2}
Br(B_{q}\to\mu^+\mu^-)
=C\frac{\tau(B_{q})}{\hat B_{B_{q}}}
\frac{Y^2(x_t)}{S(x_t)} 
\Delta M_{q}, \qquad (q=s,d)
\ee
with 
\be
C={6\pi}\frac{\eta_Y^2}{\eta_B}
\left(\frac{\alpha}{4\pi\sin^2\theta_{W}}\right)^2\frac{m_\mu^2}{\mw^2}
=4.39\cdot 10^{-10}
\ee
and $S(x_t)=2.32\pm 0.07$ and $Y(x_t)=0.94\pm0.03$ being the relevant 
top mass dependent
one-loop functions.
More generally we have in CMFV models
\be\label{R2a}
\frac{Br(B_{q}\to\mu\bar\mu)}{\Delta M_{q}}
=4.4 \cdot 10^{-10} \frac{\tau(B_{q})}{\hat B_{q}} F(v),
 \qquad F(v)=\frac{Y^2(v)}{S(v)}.
\ee
Using these expressions one finds in the SM rather precise predictions
\be
Br(B_s\to \mu^+\mu^-)= (3.6\pm0.3)\cdot 10^{-9}, \qquad
Br(B_d\to \mu^+\mu^-)= (1.1\pm0.1)\cdot 10^{-10}.
\ee
These predictions should be compared to the $95\%$ C.L. upper limits from 
CDF  \cite{Aaltonen:2007kv} 
\be\label{CDFD0}
Br(B_s\to \mu^+\mu^-)\le 6\cdot 10^{-8}, \qquad
Br(B_d\to \mu^+\mu^-)\le 2 \cdot 10^{-8}.
\ee
While the bounds from D0 \cite{Abazov:2007iy} were a bit weaker, the 2009
bounds will be similar to CDF ones.
It is clear that a lot of room is still left for NP contributions.

{\bf 5.} Next it is possible to derive a very 
accurate formula for $\sin 2\beta$ that depends only on the two
$K\to\pi\nu\bar\nu$ branching 
ratios and a calculable parameter $P_c(X)$ that represents charm
contribution to the decay $\kpn$ \cite{Buchalla:1994tr}:
\begin{equation}\label{sin2bnunu}
\sin2\beta= \frac{2 r_s}{1+r_s^2}, \qquad 
r_s=\sqrt{\sigma}{\sqrt{\sigma(B_1-B_2)}-P_c(X)\over\sqrt{B_2}}\,,
\end{equation} 
where $\sigma=1/(1-\lambda^2/2)^2$ 
and we have assumed $X>0$. The corresponding formula valid also for $X<0$ 
exists \cite{Buras:2001af}.
Here we have defined the ``reduced'' branching ratios 
\begin{equation}\label{b1b2}
B_1={Br(\kpn)\over 5.27\cdot 10^{-11}}\qquad
B_2={Br(\klpn)\over 2.27\cdot 10^{-10}},
\end{equation}
with the numerical factors obtained from \cite{Mescia:2007kn}.
Reviews on $K\to\pi\nu\bar\nu$ decays can be found in 
\cite{Buras:2004uu,Isidori:2007zs}.

It should be stressed that this formula is valid for the full class of 
MFV models and $\sin 2\beta$ determined this way depends
only on two measurable branching ratios and on the function
$P_c(X)$ which is dominated by  perturbative contributions and is known with
NNLO accuracy \cite{Buras:2006gb,Brod:2008ss}.
Also the small hadronic contributions are known \cite{Isidori:2005xm}.
The theoretical uncertainties in the determination of $\sin 2\beta$
in this manner amount to at most $1\%$ and with 
the measurements of $Br(\kpn)$ and $Br(\klpn)$
with $5-10\%$ accuracy a useful determination of $\sin 2\beta$ 
should be possible.

{\bf 6.} Moreover, as in CMFV models there are no phases beyond the KM one, 
 we also expect that \cite{Buchalla:1994tr,Buras:2001af}
\begin{equation}\label{R7}
(\sin 2\beta)_{\pi\nu\bar\nu}=(\sin 2\beta)_{J/\psi K_S}, 
\qquad
(\sin 2\beta)_{\phi K_S}\approx (\sin 2\beta)_{J/\psi K_S}.
\end{equation}
with the accuracy of the last relation at the level of a few percent 
\cite{Grossman:1997gr,Beneke:2005pu}.
The confirmation of these two relations is a very important test for the 
MFV idea. Indeed, in $K\to\pi\nu\bar\nu$ the phase $\beta$ originates in 
the $Z^0$ penguin diagram, whereas in the case of the $B\to \psi K_S$ CP 
asymmetry from 
the $B^0_d-\bar B^0_d$ box diagram. The CP asymmetry in
$B_d\to\phi K_S$  originates also in $B^0_d-\bar B^0_d$ box diagram 
but the second relation in (\ref{R7}) could be spoiled by new physics 
contributions in the decay amplitude for $B\to \phi K_S$ that is
non-vanishing only at the one loop level.  We will return to the second
relation in (\ref{R7}) below. The first relation being very clean is
the second golden relation of CMFV.

An important consequence of (\ref{sin2bnunu}) and (\ref{R7}) is the following
one. 
For a given $(\sin 2\beta)_{\psi K_S}$ 
and $Br(\kpn)$ 
only two values of 
$Br(\klpn)$, corresponding to two signs of $X$, are possible 
in the full class of CMFV models, independently of any new parameters 
present in these models \cite{Buras:2001af}. 
Consequently, measuring $Br(\klpn)$ will 
either select one of these two possible values or rule out all CMFV models.

{\bf 7.} Last but certainly not least the CP asymmetry $S_{\psi\phi}$
         is predicted in the CMFV and also MFV models to be small:
          $S_{\psi\phi}=0.04$.
         This leaves a lot of room for non-MFV contributions. We will return
         to this important observable below.

 The stringent correlations between various observables in the CMFV framework
 listed above imply rather strong upper bounds on the branching ratios of 
rare $K$ and 
 $B$ decays \cite{Bobeth:2005ck}. Including the additional constraint from 
 $Z\to b\bar b$ makes these bounds even stronger \cite{Haisch:2007ia}.
 Typically departures from SM expectations by more than $50\%$ are not allowed
 in this framework any longer.

\section{Theory of MFV}
When one considers MFV at large, new operators that are strongly suppressed
in the SM and CMFV models enter the game modifying or even removing
 the correlations present in the latter models. Consequently larger NP
effects are  allowed in these models. A recent model independent analysis of
$\Delta F=1$ processes \cite{Hurth:2008jc} finds the most interesting effects
of this type in the $B_{s,d}\to \mu^+\mu^-$ decays, 
where the presence of scalar 
operators can enhance their branching ratios up to the existing experimental
bounds in (\ref{CDFD0}). Also the branching ratios for 
$K\to\pi\nu\bar\nu$ decays and the 
forward-backward asymmetry in $B\to K^*l^+l^-$ can be sizably modified.
The corresponding analysis for $\Delta F=2$ processes in 
\cite{Bona:2005eu,Bona:2007vi} 
shows that in this sector this framework is already rather constrained.

It should be emphasized that it will be a great challenge to prove that 
MFV is the whole story
at low energies. It will be much easier to disprove
it through the violation of the correlations listed above and through
non-SM CP-violating effects, provided such non-MFV effects are really 
present. Also the pattern of the CKM matrix with strongly
suppressed couplings of the third generation of quarks to the first two is
very characteristic for the MFV models and could even be tested at the LHC.
Similar tests of MFV at the LHC have been discussed in the literature
\cite{Grossman:2007bd,Hiller:2008wp}. There further references to collider
tests of MFV can be found.

\section{Status of CMFV and MFV}
Presently both CMFV and MFV are in a good shape even if some signals of
departures from these simplest frameworks exist. We will list them
in the next section. 

The parameters of the CKM matrix have been already strongly constrained 
through the measurements of tree level processes
and of loop induced observables like the $B^0_{d,s}-\bar B^0_{d,s}$ mixing
mass diferences $\Delta M_{d,s}$, $\varepsilon_K$ in $K_L\to\pi\pi$
decays and very importantly through the mixing induced CP asymmetry 
$S_{\psi K_S}$. We have
\be
\vus=0.2255\pm0.0010, \qquad \vcb=(41.2\pm1.1)\cdot 10^{-3}, \qquad
\beta=\beta_{\psi K_S}=(21.1\pm0.9)^\circ,
\ee
where the last number follows from \cite{Barberio:2007cr}
\be
\sin 2\beta=0.670\pm0.023.
\ee
Also the following ratio is well known \cite{Barberio:2007cr}:
\be\label{vtdvtse}
\left|\frac{V_{td}}{V_{ts}}\right|=0.207\pm0.001\pm0.006,
\ee
with the given errors being experimental and theoretical errors, respectively.

It should be mentioned that the value for $\vcb$ quoted above results
from inclusive and exclusive decays that are not fully consistent with
each other. Typically the values resulting from exclusive decays are 
below $40\cdot 10^{-3}$. It would be important to clarify this difference
which has been with us already for many years. Hopefully, the future Super B
facilities in Italy and Japan and new theoretical ideas 
will provide more precise values. The
ratio in (\ref{vtdvtse}) could still be polluted by new physics. To
get the true value one would need precise values of $|V_{ub}|$ and $\gamma$
as discussed below.

Next, the angle $\alpha$ is already well determined from 
$B_d\to\varrho\varrho$ and $B_d\to\varrho\pi$ decays 
\cite{Barberio:2007cr}:
\be
\alpha=(91.4\pm 4.6)^\circ.
\ee
A specific analysis employing the mixing induced CP asymmetries 
$S_{\psi K_S}$, $S_{\varrho\varrho}$
 and the QCDF approach finds \cite{Bartsch:2008ps} $\alpha=(87\pm6)^\circ$. 
Summaries of other determinations of $\alpha$ exist
\cite{Buchalla:2008jp}.

On the other hand the status of $|V_{ub}|$ and $\gamma$ from tree level decays
is not as impressive:
\begin{displaymath}\label{vub}
|V_{ub}| =\left\{
\begin{array}{ll}
(4.0\pm0.3)\cdot 10^{-3} \,  & \mbox{(inclusive),}\\
(3.6\pm0.4)\cdot 10^{-3} \,  & \mbox{(exclusive).}
\end{array}
\right.
\end{displaymath}

\begin{displaymath}\label{gamma}
 \gamma=\left\{
\begin{array}{ll}
(78\pm12)^\circ \,  & \mbox{(UTfit),}\\
(76^{+16}_{-23})^\circ \,  & \mbox{(CKMfitter).}
\end{array}
\right.
\end{displaymath}

It is very important to precisely measure  $|V_{ub}|$ and $\gamma$ in 
the future as they determine the so-called reference UT \cite{Goto:1995hj}, 
that is free from NP
pollution. The angle $\gamma$ should be measured at LHCb to better than
$5^\circ$ accuracy in the
first half of the next decade. A precise measurement of 
$|V_{ub}|$ will require better understanding of hadronic uncertainties
and can only be performed at a Super-B facilities in Italy and Japan.

\begin{figure}[htbp]
\begin{center}
\includegraphics[width=2.7in]{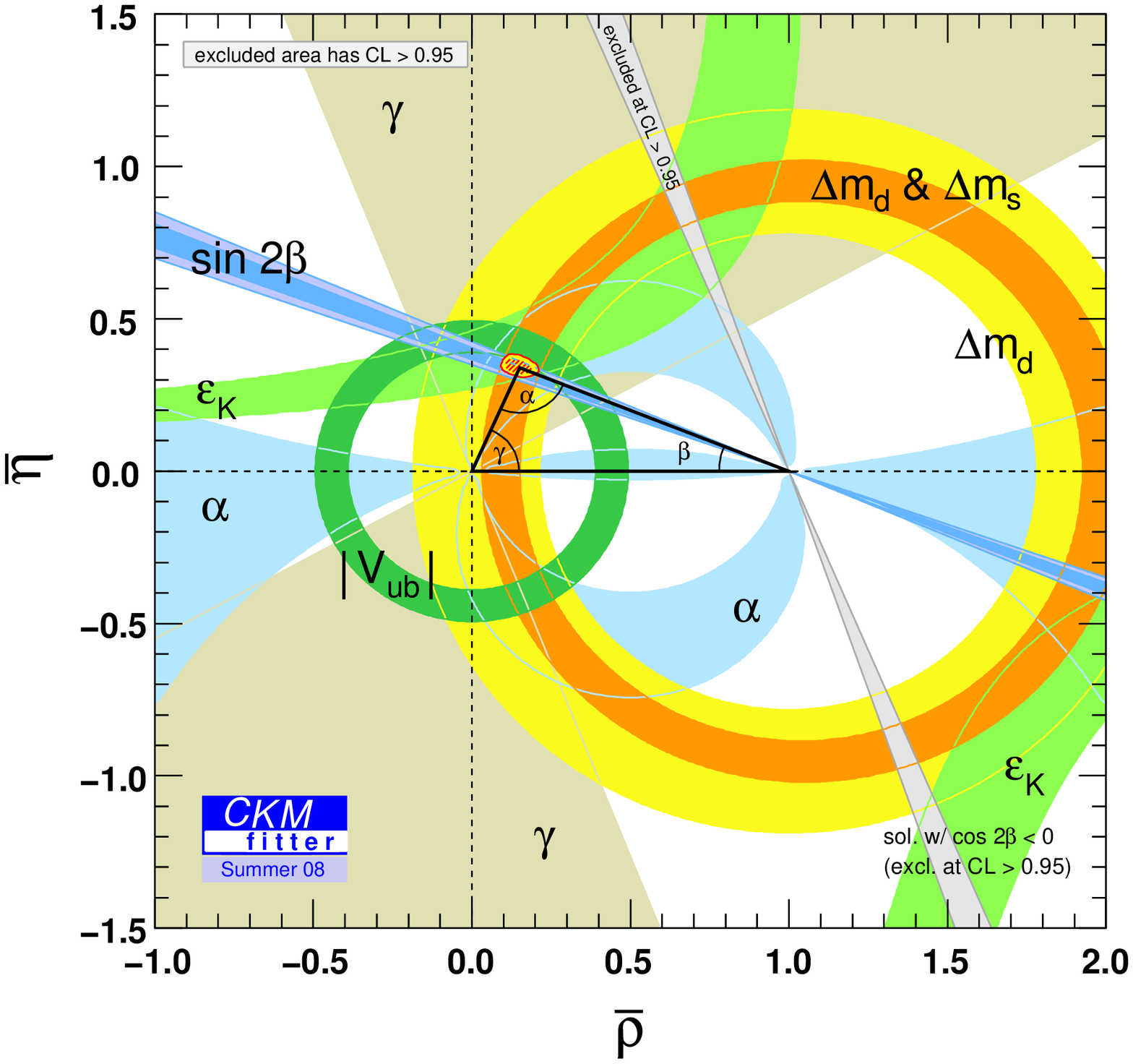}
\includegraphics[width=2.7in]{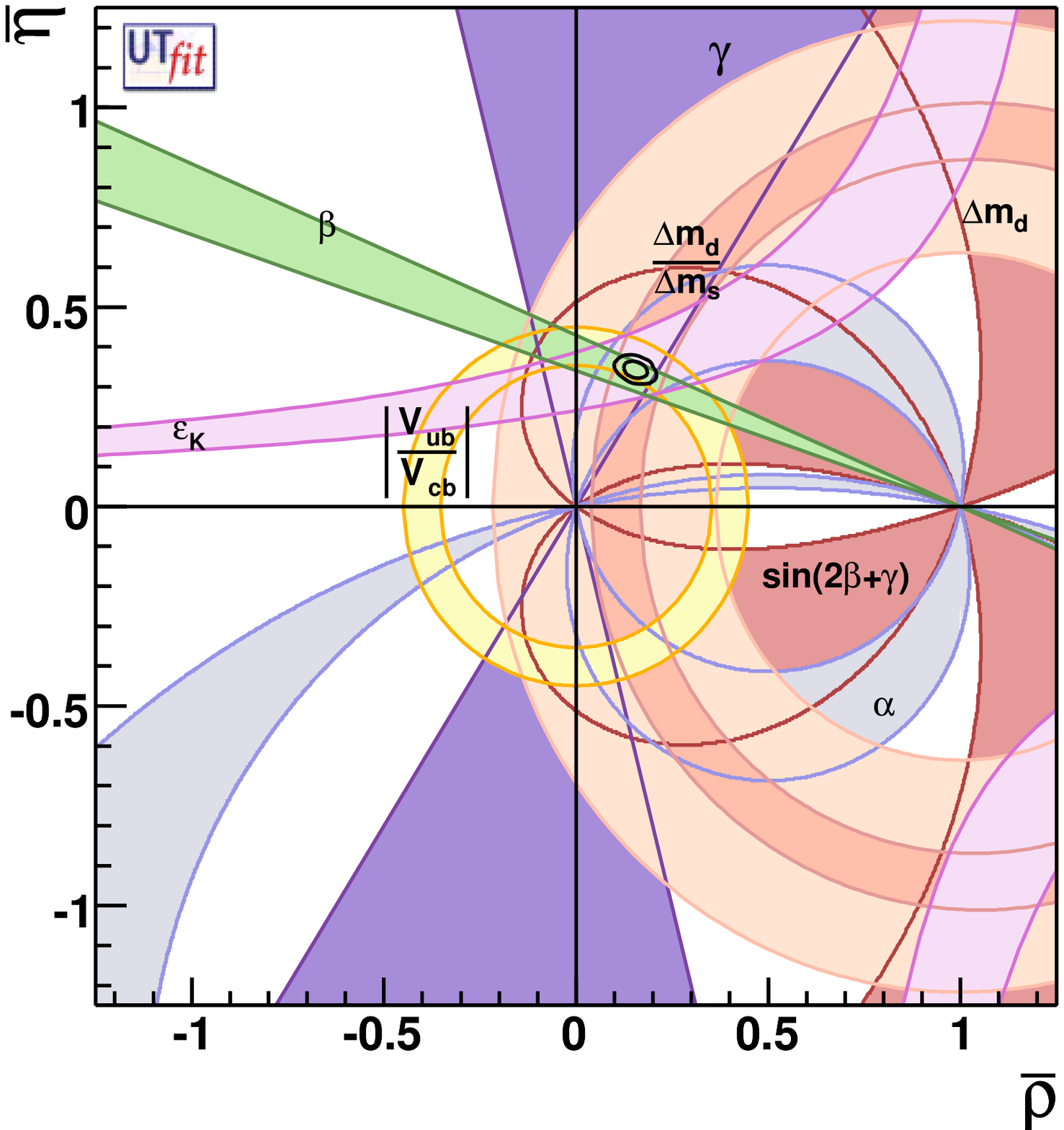}
\end{center}
\caption{\label{UT12} Unitarity triangle fits by 
CKMfitter \cite{Charles:2004jd} (left) and 
UTfit \cite{Bona:2006sa} (right) 
collaborations in 2009.}
\end{figure}

The unitarity triangle fits are shown in Fig.~\ref{UT12}.
The parameters $\bar\varrho$ and $\bar\eta$ corresponding to these plots
are given as follows

\begin{displaymath}\label{rhobar}
 \bar\varrho=\left\{
\begin{array}{ll}
0.156\pm0.020 \,  & \mbox{(UTfit),}\\
0.139^{+0.025}_{-0.027} \,  & \mbox{(CKMfitter).}
\end{array}
\right.
\end{displaymath}

\begin{displaymath}\label{etabar}
 \bar\eta=\left\{
\begin{array}{ll}
0.342\pm0.013 \,  & \mbox{(UTfit),}\\
0.341^{+0.016}_{-0.015} \,  & \mbox{(CKMfitter).}
\end{array}
\right.
\end{displaymath}

We emphasize that these results do not include new corrections to 
$\varepsilon_K$ discussed below so that the tension between the values
of $\varepsilon_K$ and $S_{\psi K_S}$ within the SM is not visible 
in these plots.

\section{Puzzles}
The CMFV and MFV frameworks appear at first sight compatible with all 
the existing data.
On the other hand, a closer look at several CP violating observables 
indicates
that the CKM phase might not be sufficient to  simultaneously describe
CP violation in $K$, $B_d$ and $B_s$ decays. In particular:
\begin{itemize}
\item
Some modes dominated by penguin diagrams, such as $B\to (\phi, \eta^{\prime}, \pi^{0}, \omega)K_{S}$ that, similarly to the golden mode $B\to \psi K_S$, allow the
determination of $\sin 2\beta$, result in $\sin 2\beta$ visibly lower than
$(\sin 2\beta)_{\psi K_S}= 0.670\pm 0.023$~\cite{Barberio:2007cr} from $B\to \psi K_S$. 
For the theoretical
cleanest modes it is experimentally found that 
$(\sin 2\beta)_{\phi K_S}= 0.44\pm 0.17$
and $(\sin 2\beta)_{\eta^{\prime} K_S}= 0.59\pm 0.07$~\cite{Barberio:2007cr}.
\item
 With the decreased value of the non-perturbative parameter $\hat{B}_{K}$ from lattice
simulations \cite{Allton:2008pn} and the inclusion of additional negative contributions to $\epsilon_K$
that were neglected in the past \cite{Buras:2008nn,Buras:2009pj}, CP violation in the $B_d-\overline{B}_d$ system,
represented by $(\sin 2\beta)_{\psi K_S}$, appears insufficient to describe 
the experimental
value of $\epsilon_K$ within the SM if the $\Delta M_d/\Delta M_s$ constraint is taken into
account. In fact we find \cite{Buras:2008nn,Buras:2009pj}
\be
\frac{|\varepsilon_K|_{\rm SM}}{|\varepsilon_K|_{\rm exp}}= 0.80\pm0.11~.
\ee
If confirmed by more precise values of $\hat B_K$ and 
more precise values of the CKM parameters, in particular $\vcb$, which enters 
roughly as $\vcb^4$ in $\varepsilon_{\rm K}$, this could
signal new physics in $\varepsilon_{\rm K}$. 
Alternatively, no new physics in
$\varepsilon_{\rm K}$ would imply $\sin 2\beta=0.88\pm 0.11$ 
\cite{Lunghi:2008aa,Buras:2008nn}.
 This could
only be made consistent with the measured value of
 $S_{\psi K_{\rm S}}$ by introducing a new phase $\phi_{\rm new}$
 in $B_d^0-\bar
  B_d^0$ mixing.
Other
possibilities are discussed in \cite{Buras:2008nn,Buras:2009pj}.
\item
There are some hints for the very clean asymmetry $S_{\psi\phi}$ to be significantly 
larger than the SM value $S_{\psi\phi}\approx 0.04$ 
\cite{Abazov:2007zj,Abazov:2007tx,Abazov:2008fj,Brooijmans:2008nt}.
Theoretical papers related to these results can be
  found
in \cite{Lenz:2006hd,Bona:2008jn,Lenz:2008dp}. 
\item
The rather large difference
in the direct CP asymmetries $A_{CP}(B^{-}\to K^{-}\pi^{0})$ and
$A_{CP}(\overline{B}^{0}\to K^{-}\pi^{+})$ observed by the Belle and BaBar 
collaborations has not been  expected but it could be due to
our
insufficient understanding of hadronic effects rather than NP. 
Similar comments apply to certain puzzles 
in $B\to\pi K$ decays \cite{Buras:2004ub} which
represent additional tensions that decreased with time but did not
fully disappear \cite{Baek:2009pa}. 
\item
Finally, there is the muon anomalous magnetic moment anomaly. Most recent analyses converge
towards a $3\sigma$ discrepancy in the $10^{-9}$ 
range~\cite{Passera:2004bj,Passera:2005mx}:
$\Delta a_{\mu}\!=\!a_{\mu}^{\rm exp}\!-\!a_{\mu}^{\rm SM}\approx(3\pm 1)\times 10^{-9}$
where $a_{\mu}\!=\!(g-2)_{\mu}/2$.
Despite substantial progress both on the 
experimental~\cite{Bennett:2002jb,Bennett:2004pv} and on
the theoretical sides,
the situation is not completely clear yet. However, the possibility that the
present discrepancy
may arise from errors in the determination of the hadronic leading-order contribution to
$\Delta a_{\mu}$ seems to be unlikely, as stressed in 
\cite{Passera:2008jk}. Recent reviews can be found in 
\cite{Passera:2008hj,Jegerlehner:2009ry}.
\end{itemize}

\section{Beyond CMFV and MFV}
\subsection{Most Popular Non-MFV Extensions of the SM}
Having possible signals of non-MFV interactions at hand, let us consider
briefly the most popular non-MFV extentions of the SM that are connected
with the hierarchy problem related to  quadratic divergences in 
the Higgs 
mass and the disparity of the electroweak, GUT and Planck scales. 
The three most promising and
 most popular 
directions which aim to solve at least some of these problems are as follows:

{\bf a) Supersymmetry.} In this approach the cancellation of divergences in $m_H$ is achieved with 
the help of 
new particles of different spin-statistics than the SM particles:
supersymmetric particles.
 For this approach to 
work, these new particles should have masses well below 1 TeV, 
otherwise  fine 
tuning of parameters cannot be avoided. As none of the 
supersymmetric particles 
has been seen so far, the MSSM became a rather fine tuned scenario even if much
less than the SM in the presence of the GUT and Planck scales. 
One of the important predictions of 
the 
simplest realization of this scenario, the MSSM with R-parity, 
is  light Higgs with 
$m_H\le 130\gev$ and one of its virtues is its perturbativity up to the GUT
scales. 
The ugly feature of the MSSM is a large number of parameters residing 
dominantly in the soft sector
that has to be introduced in the process of supersymmetry breaking. 
Constrained 
versions of the MSSM can reduce the number of parameters significantly. 
The same is true in the case of the MSSM with MFV. An excellent introduction
to the MSSM can
be found in \cite{Martin:1997ns}.

 Concerning the FCNC processes let us recall that  in addition to a 
light Higgs, 
squarks, sleptons, gluinos, charginos and
 neutralinos, also charged Higgs particles $H^{\pm}$ and additional 
 neutral scalars are present in this framework. 
All these particles can contribute to FCNC transitions through box and 
penguin diagrams. 
New sources of flavour
and CP violation come from the misalignement of quark and squark mass 
matrices
 and similar new flavour and CP-violating effects are present in the lepton
sector. Some of these effects can be strongly enhanced at large $\tan\beta$
and the corresponding observables provide stringent constraints on
the parameters of the MSMM.
In particular $B_s\to \mu^+\mu^-$ can be enhanced up to its experimental
upper bound, branching ratios for $K\to\pi\nu\bar\nu$ can be much larger than
their SM values and the CP asymmetry $S_{\phi\psi}$ can also strongly
deviate from the tiny SM value.

There is a very rich literature on FCNC processes in general supersymmetric
models and the large number of parameters present in these models, while 
allowing to make numerous phenomenological analyses, precludes often 
 clear cut conclusions.

The flavour blind MSSM (FBMSSM) scenario \cite{Baek:1998yn,Baek:1999qy,Bartl:2001wc,Ellis:2007kb,Altmannshofer:2008hc} having new 
complex phases that are flavour conserving
belongs actually to the class of MFV models but as the functions $F_i$
become complex quantities we mention this model here. 
The FBMSSM has fewer parameters than the general MSSM(GMSSM)
 implying striking correlations
between various observables. In particular the desire to explain the 
suppression of $S_{\phi K_S}$ below $S_{\psi K_S}$ implies a direct CP
asymmetry in $B\to X_s\gamma$ that is by one order of magnitude larger
than its SM value. Also $d_n$, the electric dipole moment of the neutron,
is found to be as high as $10^{-28}{\rm e.cm}$ for the same 
reason. The FBMSSM belongs to the more general models with MFV discussed
recently \cite{Kagan:2009bn}.

{\bf b)	Little Higgs Models.}
In this approach the cancellation of divergences in $ m_H$ is achieved with 
the help of new particles of the same spin-statistics. 
Basically the SM Higgs is kept light 
because it is a pseudo-Goldstone boson of a spontaneously broken global 
symmetry. 
Thus the Higgs is protected from acquiring a large mass by a global symmetry, 
although in order to achieve this the weak gauge group has to be extended and 
the Higgs mass generation properly arranged 
({\it collective symmetry breaking}). The 
dynamical origin of the global symmetry in question and 
the physics behind its 
breakdown are not specified. But in analogy to QCD one could imagine 
a new strong 
force at scales $\ord(10-20\tev)$ between  new very heavy fermions that bind 
together to produce the SM Higgs. In this scenario the SM Higgs is 
analogous to 
the pion. At 
scales well below $ 5\tev$ the Higgs is considered as an elementary particle but at
$20\tev$  its 
composite structure should be seen.  Possibly at these high scales one will
have to cope with non-perturbative strong dynamics and an unknown ultraviolet 
completion with some impact on 
low energy predictions of 
Little Higgs models has to be specified. Concrete perturbative 
completions, albeit very complicated, have been found 
\cite{Batra:2004ah,Csaki:2008se}.
The advantage of these models, relative to supersymmetry, is a 
much smaller number of free parameters but the disadvantage is the presence
of new matter and new interactions on the way to the GUT scale so that  
Grand Unification
in this framework is rather unlikely. Excellent reviews can be found in
\cite{Schmaltz:2005ky,Perelstein:2005ka}.

Concerning the FCNC processes let us recall that in contrast to the MSSM, 
new heavy gauge bosons $W_H^\pm$, $Z_H$ and
 $A_H$  in the case of the so-called littlest Higgs model 
without \cite{Arkani-Hamed:2002qy} and with T-parity 
\cite{Cheng:2003ju,Cheng:2004yc} are present. Restricting our 
discussion to the model with T-parity (LHT), the masses of $W_H^\pm$ and
 $Z_H$ are typically $\ord(700\gev)$. $A_H$ is significantly lighter with
 a mass of a few hundred GeV and, being the lightest particle 
with odd T-parity, it can play the role of a dark matter candidate. 
Concerning the 
fermion sector, there is a new heavy $T$-quark necessary to cancel 
the quadratic 
divergent contribution of the ordinary top quark to $m_H$ and 
a copy of all SM
quarks and leptons, required by T-parity. These mirror quarks and mirror 
leptons 
interact with SM particles through the exchange of
$W_H^\pm$, $Z_H$ and  $A_H$ gauge bosons that in turn implies 
 new flavour and CP-violating contributions to decay amplitudes. 
These new contributions are 
governed by  new mixing matrices
in the quark and lepton sectors which
 can have a structure very different from the CKM and PMNS matrices.
The mirror quarks and leptons can have masses typically in the range 
500-1500 GeV and could be discovered at the LHC. 
Their impact on
FCNC processes can be sometimes spectacular. Reviews on flavour physics in 
the LHT model can be found in 
\cite{Blanke:2007ww,Duling:2007sf} and selected papers containing details
of the pattern of flavour violation in these models can be found in
\cite{Blanke:2006sb,Blanke:2006eb,Blanke:2007db,Goto:2008fj,delAguila:2008zu,Blanke:2009pq}.
In particular the asymmetry $S_{\psi\phi}$ can be much larger than its
SM value \cite{Blanke:2008ac}, the rare decays $K\to\pi\nu\bar\nu$ 
can be strongly enhanced \cite{Blanke:2006eb,Goto:2008fj}  and the effects in lepton flavour violating
decays like $\mu\to e\gamma$ can be very large 
\cite{Blanke:2007db,delAguila:2008zu}. 

Recently also CP violation in  $D^0-\bar D^0$ mixing has been analyzed in
this framework \cite{Bigi:2009df}. Observable effects at a level well beyond anything
possible with CKM dynamics have been identified. Comparisons with CP violation
in $K$ and $B$ systems should offer an excellent test of this NP scenario and
reveal the specific pattern of flavour and CP violation in the $D^0-\bar D^0$
system predicted by this model.

{\bf c)	Extra Space Dimensions.}
When the number of space dimensions is increased,
 new solutions to the hierarchy 
problems are possible. Most 
ambitious proposals are models with a warped extra dimension first proposed
by Randall and Sandrum (RS)  \cite{Randall:1999ee} which provide a geometrical
explanation of the
hierarchy  between the Planck scale and the EW scale. Moreover, when the SM
fields, except for the Higgs field, are
allowed to propagate in the bulk 
\cite{Gherghetta:2000qt,Chang:1999nh,Grossman:1999ra}, 
these models naturally generate the
hierarchies in the fermion masses and mixing angles 
\cite{Grossman:1999ra,Gherghetta:2000qt} while simultaneously 
suppressing FCNC transitions with the 
help of the so-called RS-GIM mechanism ~\cite{Huber:2003tu,Agashe:2004cp}.
 Yet, in these models FCNC processes appear
already at tree level 
\cite{Burdman:2003nt,Huber:2003tu,Agashe:2004cp,Csaki:2008zd} and in the
case of $\varepsilon_K$ which receives tree level KK gluon contributions
 some fine-tuning
of parameters in the flavour sector is necessary in order to achieve 
consistency with the data for KK scales in the reach of LHC 
\cite{Csaki:2008zd,Blanke:2008zb}.

Moreover, to avoid problems
with electroweak precision tests (EWPT) and FCNC processes, the gauge group 
is
generally larger than the SM gauge group and similarly to the LHT model
new heavy gauge bosons are present. However, even in models with custodial
symmetries \cite{Agashe:2003zs,Csaki:2003zu,Agashe:2006at}, these gauge bosons must be sufficiently heavy ($2-3\tev$) 
in order to be consistent with EWPT. 

In the case of rare $K$ and $B$ decays
the RS-GIM mechanism 
\cite{Huber:2003tu,Agashe:2004cp} combined with additional custodial
protection of  flavour violating $Z$ couplings 
\cite{Blanke:2008zb,Blanke:2008yr}
allows  to achieve 
agreement with
existing data without a considerable fine tuning of parameters
\cite{Blanke:2008zb,Blanke:2008yr} and still produce interesting 
effects in observables that should be measured in the coming years. 
Most recent reviews on the latter
work can be found in
\cite{Duling:2009sf,Gori:2009tr}. One finds in these models a clear pattern 
of flavour violation: large effects in $\Delta F=2$ transitions and
rare $K$ decays but small effects in rare $B$ decays except for
$B\to X_s\gamma$ \cite{Agashe:2008uz}. 
However, simultaneous large effects in $\Delta F=2$ 
processes and rare $K$ decays are rather unlikely.
Large effects are also found in $\mu\to e\gamma$ \cite{Agashe:2006iy}
and electric dipole moments. A detailed presentation of a particular
model with custodial protection including Feynman rules 
exists \cite{Albrecht:2009xr}. Very recently possible flavour protections
in warped Higgsless models have been presented \cite{Csaki:2009bb}. On
the other hand various aspects of flavour physics in a model without
custodial protections have been discussed by the Mainz group 
\cite{Casagrande:2008hr,Bauer:2008xb}.

\subsection{The Flavour Matrix}
\begin{figure}
\begin{center}
\includegraphics[width=3.2in,angle=270]{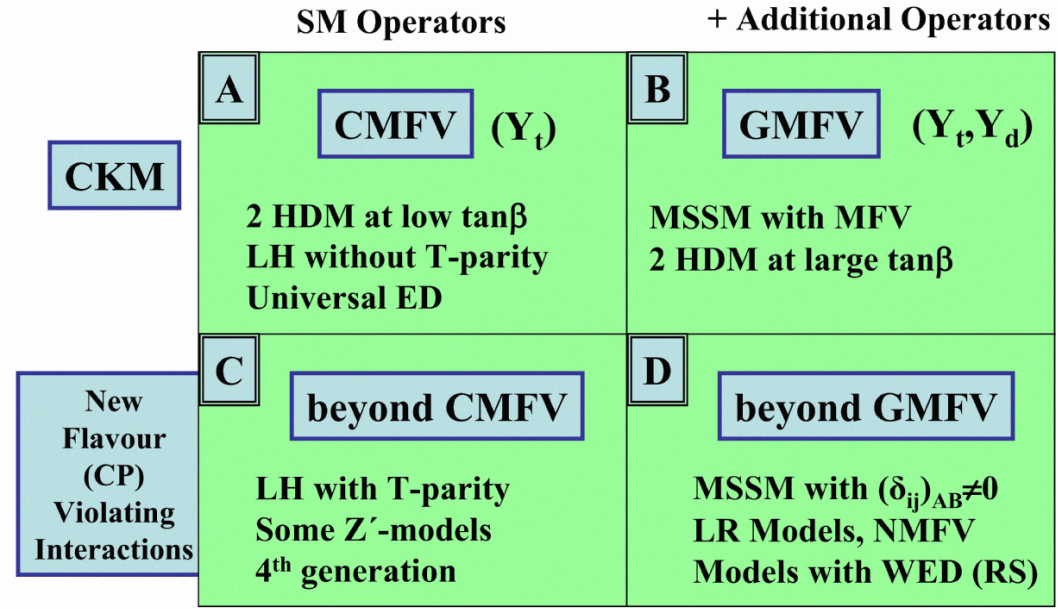}
\end{center}
\caption{\label{fig:matrix}The Flavour Matrix}
\vspace*{-0.2cm}
\end{figure}

After the discussion of CMFV, MFV and various non-MFV
extensions of the SM let us compare them from the point of view of
the presence of new operators and/or new sources of flavour violation
with respect to the SM. Our discussion of Section~4  results  in
a $2\times 2$ matrix  shown in Fig.\ref{fig:matrix}.
Let us briefly describe the four
entries in this matrix.

The element (1,1) or the class A represents the models with 
CMFV discussed in detail in Section~4.
We have  seen above  that this class of models does not allow for 
large deviations from the SM predictions.

The elements (1,1) and (1,2) or classes A and B taken together, 
the upper row of the flavour matrix, 
represent the class of models with MFV at large that we discussed 
briefly in Section~5.
Basically the new effect in the
(1,2) entry  relative to 
(1,1) alone is the appearance of new operators 
with different Dirac structures that 
are strongly suppressed in the CMFV framework but can be 
enhanced if $\tan\beta$ is large or 
equivalently if $Y_d$ cannot be neglected. 
The presence of new
operators,  in particular scalar 
operators, allows to lift the helicity suppression of certain rare decays 
like $B_s\to\mu^+\mu^-$, 
resulting in very different predictions than found in CMFV models.

A very interesting class of models is the one represented by the 
(2,1) entry or the class C. Relatively to CMFV it contains new flavour 
violating interactions, in
particular new complex phases, forecasting novel
 CP-violating effects that may significantly differ from 
those present in the CMFV class. 
As there are no new operators relatively to the SM ones, 
no new $B_i$-factors and 
consequently no new non-perturbative uncertainties 
relative to CMFV models are 
present. 
Therefore predictions of models belonging to the (2,1) entry suffer generally 
from smaller non-perturbative uncertainties 
than models represented by the second 
column in the flavour matrix in Fig.~3.

When discussing the models in (2,1), it is important to distinguish 
between models in 
which NP couples dominantly to the third generation of 
quarks, basically 
the top quark, and models where there is a new sector of fermions that can 
communicate with the SM fermions with the help of new gauge interactions. 
Phenomenological approaches with enhanced Z-penguins 
\cite{Buras:1998ed,Buras:1999da,Buras:2004ub}, some special
$Z^\prime$-models \cite{Langacker:2000ju} and the fourth generation models
\cite{Hou:2008di,Soni:2008bc,Bobrowski:2009ng}
belong to the first subclass of (2,1), while the
LHT model belongs to the second subclass.

Finally there is the most complicated class of models represented 
by the (2,2) entry or the class D
in which not only new flavour violating effects but also new operators 
are relevant. 
The MSSM with flavour violation coming from the squark sector and RS 
models
are likely to be  the most 
prominent members of this class of models.
In  the RS models FCNC transitions take place already at tree level and the
pattern of flavour violation in these models generally differs from the LHT
model and the MSSM \cite{Blanke:2008yr}.
 NMFV \cite{Agashe:2005hk} and left-right symmetric models belong also 
to this class. A spurion 
technology to classify these models has been developed by Feldmann and
Mannel \cite{Feldmann:2006jk}.

\section{Outlook}
The frameworks of CMFV and MFV in which the CKM matrix is the only source
of flavour and CP violation in the quark sector could turn out to be  
the correct
description of the full class of flavour violating processes. On the other
hand many new branching ratios and many CP-violating observables will be
measured in the coming years with high precision. Therefore we should be
prepared for surprises and the studies of various extentions of the SM with
non-MFV interactions indicate that these surprises could still be spectacular.

In particular:
\begin{itemize}
\item
The measurement of $S_{\psi\phi}$ with a value above $0.2$ would signal a
clear
violation of both CMFV and MFV.
\item
The measurement of $Br(B_s\to\mu^+\mu^-)$ above $5\cdot 10^{-9}$ would 
be inconsistent with CMFV, signalling the presence of new (scalar)
operators, but would still be consistent with MFV at large.
\item
The measurement of electric dipole moment of the neutron at the level of 
$10^{-27}{\rm e.cm}$
would be inconsistent with the CKM picture by several orders of magnitude.
\end{itemize}

These are just three out of many possible prominent signals that would
definitely imply NP beyond the SM and beyond the CKM picture. Later
$K\to\pi\nu\bar\nu$ and $K_L\to \pi^0l^+l^-$  decays 
could be used to obtain a deeper
insight into the flavour structure at very short distance scales. Correlations
between $B$ and $K$ decays will play an important role in this context. 
Also CP violation in $D^0-\bar D^0$ mixing 
\cite{Blum:2009sk,Grossman:2009mn,Bigi:2009df} and searches for
FCNC processes in the up-quark sector 
at the LHC will be very helpful in this respect. Nice recent reviews of the
physics of CP violation and flavour violation 
have been presented by Fleischer \cite{Fleischer:2008uj}
 and Nierste \cite{Nierste:2009wg},  where
further references can be found.

If no spectacular deviations from the SM will be observed, there is still
a multitude of correlations between various observables, in particular 
the two golden relations discussed in Section~4 that eventually could
tell us how precise the CKM picture is. In any case the next decade should be
very exciting for flavour physics resulting hopefully in new Nobel Prizes
for advances in this fascinating field.

\section*{Acknowledgements}
I would like to thank all my collaborators for a wonderful
time we spent together exploring different avenues beyond the Standard
Model. Special thanks go to Monika Blanke and Bj\"orn Duling for comments
on the manuscript.
This research was partially supported by the Deutsche
Forschungsgemeinschaft (DFG) under contract BU 706/2-1, the DFG Cluster of
Excellence `Origin and Structure of the Universe' and by the German
Bundesministerium f{\"u}r Bildung und Forschung under contract 05HT6WOA.

%

  


\addcontentsline{toc}{chapter}{References}
\providecommand{\href}[2]{#2}\begingroup\raggedright\endgroup
\end{document}